\documentclass[preprint, review,authoryear,12pt]{elsarticle}

\usepackage{tipa}
\usepackage{color}
\usepackage{graphicx}

\usepackage{amssymb}

\journal{Icarus}

\begin{document}

\begin{frontmatter}



\title{Survival of water ice in Jupiter Trojans}


\author[uc,esa]{Aur\'elie Guilbert-Lepoutre\corref{cor}}
\cortext[cor]{\ead{aguilber@rssd.esa.int}}
\address[uc]{UCLA - Department of Earth and Space Sciences, 90095 Los Angeles, USA}
\address[esa]{European Space Agency - ESTEC, 2201 AZ Noordwijk, Netherlands}

\begin{abstract}
Jupiter Trojans appear to be a key population of small bodies to study and test the models of the solar system formation and evolution. Because understanding the evolution of Trojans can bring strong and unique constraints on the origins of our planetary system, a significant observational effort has been undertaken to unveil their physical characteristics. 
The data gathered so far are consistent with Trojans having volatile-rich interiors (possibly water ice) and volatile-poor surfaces (fine grained silicates). Since water ice is not thermodynamically stable against sublimation at the surface of an object located at $\sim$5~AU, such layering seems consistent with past outgassing. 
In this work, we study the thermal history of Trojans after the formation of a dust mantle by possible past outgassing, so as to constrain the depth at which water ice could be stable. We find that it could have survived 100~m below the surface, even if Trojans orbited close to the Sun for $\sim$10,000~years, as suggested by the most recent dynamical models. Water ice should be found $\sim$10~m below the surface in most cases, and below 10~cm in the polar regions in some cases.
\end{abstract}


\end{frontmatter}



\section{Introduction}

Jupiter Trojans (hereafter Trojans) are asteroids trapped in the L4 and L5 Lagrangian clouds of the Jupiter-Sun system. Because they share a similar orbit as Jupiter, they have been considered as useful probes to constrain the formation of the giant planet. However, as of today, we are still mostly ignorant of the origin of this small-body population, and this topic is a matter of great debate. Trojans might have formed in the Jupiter region \citep{Mar98, Fle00}, near where they are found today. They could in this case provide important clues on the formation of Jupiter itself, as its core is believed to have formed from an aggregate of icy planetesimals \citep{Fle00}. 
Later in the dynamical history of the solar system, the Lagrangian clouds might have been populated by objects formed in more distant regions, in which case the current population could include bodies formed over a large range of heliocentric distances ($\sim$5-30~AU). Trojans could have been captured either during a chaotic \citep{Mor05, Nes13} or a smooth migration of the giant planets \citep{Lyk09}. They might represent the physical properties of their more distant and less accessible potential parent bodies, the Kuiper Belt Objects (KBOs). In this context, understanding the origin and fate of Trojans could lead to significant advances in our understanding of the formation and evolution of the solar system. 

Trojans appear to be a key small-body population for testing models of the solar system dynamical evolution. The physical nature of Trojans, when understood, should provide crucial constraints on the temperature, pressure and chemical composition of the solar nebula at the time and place of their formation. A large observational effort has thus been undertaken to unveil the physical characteristics of these small bodies. Their surfaces display low albedos and colors very similar to those observed amongst comets \citep{Jew90}. Spectroscopic data collected over a large wavelength range have revealed featureless spectra in the 0.4-4.0~$\mu$m range \citep{Jon90, Luu94, Dum98, Eme03, Yan07, For07, DeL10, Yan11, Eme11}, while their mid-infrared spectra display a prominent feature around 10~$\mu$m attributed to the presence of fine-grained silicates at their surface \citep{Eme06}. The inferred compositions are consistent with Trojan surfaces being made of a very porous dusty crust, possibly produced by past sublimation \citep{Eme06}. For instance, the composition of grains at the surface of Trojan (624) Hektor \citep{Ver12} appears similar to the composition of cometary grains \citep{Lis06, Bru11}. \citet{Eme11} reported a possible bimodality in the near-infrared spectral slopes, consistent with previous trends reported among visible data \citep[see][and ref. therein]{For07}. The authors argue that the two spectral groups could represent objects with different intrinsic compositions, due to different formation locations, with the reddest spectral group formed in the outer solar system, and the other objects formed in the Jupiter region. Density was determined for a few objects: it ranges from  0.8~g~cm$^{-3}$ for (617)~Patroclus \citep{Mar06} to 2.5~g~cm$^{-3}$ for (624)~Hektor \citep{Lac07}, although a recent measurement from \citet{Mar13} indicates a density of 1.0~g~cm$^{-3}$ for this object. This suggests that Trojans have a high volatile content, a high porosity, or both. Finally, the distribution of Trojan rotation periods may indicate possible past outgassing \citep{Mot12}. 

Trojans could be considered as dead or dormant comets. All data gathered so far are indeed consistent with Trojans being volatile-rich objects, with surfaces made of a porous dusty crust, possibly produced by past outgassing. However, no ice has ever been reported on Trojans, nor any coma or outgassing ever detected. Did any ice, if ever present in Trojans, survive their possible cometary past? Could it be buried in the deep interior, or more interestingly in subsurface layers? In this work, we investigate the survival of water ice inside Trojans, regardless of their early thermal evolution, or their specific past cometary activity. We use a three-dimensional model to compute the thermal evolution of Trojans as a function of albedo, obliquity, rotation period and thermal inertia, in order to provide constraints on the depth at which water ice might have survived.

\section{Thermal evolution modeling of Trojans}

\subsection{Assumption on the internal composition}
The formation of a dusty mantle at the surface of comets was first studied by \citet{Bri79}. The idea behind this process is the following: as ice evaporates, the gas carries dust particles with it. While the smaller, lighter particles are entrained and can escape the body, the larger particles accumulate on the surface, eventually creating a porous crust. In time, the insulating effect of this mantle can completely quench any sublimation \citep{Pri88,Gru93}. The realistic modeling of dust mantling, and of gas flow through such a mantle, is however extremely complex \citep{Hue06}. In this work, we do not study this phase of thermo-physical evolution. 
We consider as an approximation that Trojans sustained cometary activity in the past, which resulted in the formation of an insulating crust at their surface. In our model, the crust is formed, and ice is stable underneath, so that no sublimation needs to be accounted for. This is similar to the approach of \citet{Sch08} for Main Belt Asteroids. Our aim is to study the possibility for water ice to survive under a dust mantle, and establish the thickness of such mantle. We also check whether the conditions are met for water ice to survive at the surface of Trojans.

\paragraph{At the surface} The ice loss rate from the surface $J_{surf}$ is given by:
\begin{equation}
J_{surf} = P_S(T) \sqrt{\frac{m}{2\pi k_B T}}
\end{equation}
with $m$ [kg] the molecular weight, $k_B$ [JK$^{-1}$] the Boltzmann constant, $T$ [K] the temperature, and $P_S(T)$ [Pa] the saturation vapour pressure, which is given by the Clausius-Clapeyron equation:
\begin{equation}
\label{pression}
P_S(T) = \alpha ~ e^{-E_a/k_B T}
\end{equation}
with $\alpha=3.56 \times 10^{12}$~Pa and $E_a/k_B=6141.677$~K for water. 

\paragraph{Under the surface} The loss rate from an icy layer buried under a porous crust is driven by three main parameters:
\begin{itemize}
\item[-] the thickness of the crust, $\Delta r$ [m],
\item[-] the diffusion coefficient, $D_{Kn}$ [m$^2$s$^{-1}$], which relates to the ability of gas molecules to escape through the crust by Knudsen diffusion, 
\item[-] the temperature T [K].
\end{itemize}
The diffusion coefficient can be written as $D_{Kn}\sim\varphi ~ v_{th}$, with $v_{th}=\sqrt{\frac{8k_B T}{\pi m}}$ the mean thermal velocity, and $\varphi$ the material permeability. We can write the ice loss rate $J$ from a subsurface layer buried under the porous icy crust as (assuming an ideal gas law):
\begin{equation}\label{jint}
\begin{array}{ll}
& J = D_{Kn}~\frac{1}{\Delta r}~\frac{m}{k_B T}~P_S(T) \\
\\ 
\Longrightarrow  & J =\frac{2\varphi}{\Delta r}~\sqrt{\frac{2m}{\pi k_B T}}~P_S(T),
\end{array}
\end{equation}
which is in agreement with the expressions found by \citet{Fan84, Sch08} and \citet{Gun11} by a factor of the order of unity.

Both at and under the surface, the survival of water ice, after a potential episode of cometary activity, can be constrained by studying the temperature distribution in Trojans. At the surface, the water ice loss rate is $\sim$3$\times10^{-15}$~kg~m$^{-2}$s$^{-1}$ at 110~K. This means that water ice could survive for the age of the solar system below this temperature. We imposed the same low erosion rate in Eq.~(\ref{jint}), to directly constrain the thickness of the porous dusty crust which would be required for water ice to survive, from the temperature distribution inside the crust.

\subsection{Model for the temperature distribution}

To explore the stability of water ice, we therefore need to determine the temperature distribution. We use a numerical model of three-dimensional heat transport: the temperature distribution is computed as a function of time and orbital position, both at the surface and inside the object. In this section, we briefly described the model, and refer to \citet{Gui11} for specific details on the mathematical and numerical scheme. The model solves the heat conduction equation, assuming spherical objets. Heat is transported in two different ways: it is conducted via contacts between grains, and transferred through thermal radiation within pores. Boundary conditions are: at the center of the object, the heat flux is null; at the surface, it is given by computing the thermal balance between:
\begin{itemize}
\item[-] insolation, described by $(1-\mathcal{A})\frac{C_{\odot}}{d^2_H}\cos\xi$, with $\mathcal{A}$ the Bond albedo, $C_{\odot}$ the solar constant, $d_H$ the object's heliocentric distance, and $\xi$ the local zenith angle,
\item[-] thermal emission $\varepsilon \sigma T^4$, with $\varepsilon$ the material emissivity, $\sigma$ the Stefan-Boltzmann constant and $T$ the temperature,
\item[-] and both lateral and radial heat fluxes, driven by the thermal conductivity $\kappa$.
\end{itemize}

We consider an initial internal structure as suggested by the observational data: a core made of a mixture of porous ice and dust (50 wt\% each) uniformly distributed within the ice, and an insulating surface layer made of porous dust. 
The crust thickness is however the unknown parameter which we are trying to constrain. We thus start the calculations with an arbitrarily thick surface layer ($\Delta r_0$ of the order of 1~km), and compute the temperature distribution after 4~Gyr. By introducing this distribution in Eq.\ref{jint}, we estimate the crust thickness $\Delta r_1$, and re-adjust the numerical grid so be very fine close to this transition between the crust ad the core. This allows to minimize the discretization errors. We iterate the calculation of the temperature distribution until the depth converges (typically 2 to 3 iterations in total), so to establish the crust thickness associated to each case described in the following sections.

For the structural characteristics of the crust, we use the definition of permeability given by \citet{Hue06}: $\varphi = \frac{\psi r_p}{\tau ^2}$, with $\psi$ the porosity, $r_p$ the average pore size and $\tau$ the tortuosity. We choose two values for the porosity, 0.5 and 0.85, in order to be consistent with the thermal properties assumed below. We then use the empirical formula given by \citet{Kap97} to compute the tortuosity as a function porosity: for $\psi$=0.5 and 0.85 in the crust, $\tau$=1.45 and 1.11 respectively. We assume an average pore size $r_p$=100$\mu$m. The thermal properties of the insulating crust are described in the following section, since their effect is investigated in detail.


\subsection{Influence of parameters relevant to the study}

\paragraph{Thermal inertia}
The thermal properties of small bodies are very hard to constrain. As of today, measurements of their thermal inertia are available for a handful of Trojans, but due to their potential dynamical connection to Centaurs and comets \citep{Hor10b}, we consider measurements for those objects too. For example, \citet{Jul00} reported a thermal inertia $\Gamma$=100~J~K$^{-1}$m$^{-2}$s$^{-1/2}$ for comet 1P/Halley. However, the thermal inertia of comets is generally very low \citep{Hue06}. Measurements for Centaurs (2060)~Chiron, (8405)~Asbolus and (10199)~Chariklo point toward $\Gamma<$10~J~K$^{-1}$m$^{-2}$s$^{-1/2}$ \citep{Fer02, Gro04}. \citet{Fer03} measured $\Gamma <$14~J~K$^{-1}$m$^{-2}$s$^{-1/2}$  and $\Gamma<$30~J~K$^{-1}$m$^{-2}$s$^{-1/2}$ for Trojans (2363)~Cebriones and (3063)~Makhaon respectively. \citet{Mue10} reported an average thermal inertia of 20$\pm$5~J~K$^{-1}$m$^{-2}$s$^{-1/2}$ for Trojan (617)~Patroclus. 

Although the thermal inertia of Trojans may appear low given these measurements, \citet{Hor12} recently reported a thermal inertia between 25 and 100~J~K$^{-1}$m$^{-2}$s$^{-1/2}$ for (1173)~Anchises. We therefore consider two values for the thermal inertia of the porous dusty crust at the surface of Trojans, both low and high. This thermal inertia is considered uniform across the surface, and within the dusty crust. We follow the measurements of the thermal conductivity of porous dust made by \citet{Kra10}. For the low thermal inertia, we use a porosity $\psi$~=~0.85 and a thermal conductivity $\kappa$~=~2$\times$10$^{-3}$~Wm$^{-1}$K$^{-1}$, leading to $\Gamma \sim$~10~J~K$^{-1}$m$^{-2}$s$^{-1/2}$. For the high thermal inertia, we use $\psi$~=~0.5, $\kappa$~=~2$\times$10$^{-2}$~Wm$^{-1}$K$^{-1}$, leading to $\Gamma \sim$~100~J~K$^{-1}$m$^{-2}$s$^{-1/2}$.

\paragraph{Obliquity and rotation period}
The spin of an object is generally deduced from the variations of its brightness with time. However, the lightcurves of Trojans are not very well characterized. The available sample is small compared to the number of objects detected in the two Lagrangian swarms, and restricted to the larger objects. For example, \citet{Mot11} present the rotation period of more than 80 Trojans, which ranges from 5.7~hrs to more than 15.9~hrs. We thus consider rotation periods of 5~hrs, 10~hrs (our reference rotation period) and 15~hrs. 
In addition, it is very hard from these data to constrain the direction of the spin axis. Therefore, for the sake of simplicity, we define the obliquity $\Theta$ as the tilt of the spin axis relative to the orbital plane, and study the influence of this angle. The tilt is always pointing toward the Sun at perihelion: for $\Theta$=0$^{\circ}$, the subsolar point is always located at the equator and the poles are never illuminated, for $\Theta$=90$^{\circ}$, poles are pointing toward the Sun at perihelion and aphelion.  

\paragraph{Albedo}
\citet{Gra11} present the results of the NEOWISE survey, which observed more than 2000 Trojans, thus increasing the available sample of Trojan sizes and albedos by more than an order of magnitude. The results point toward a very homogeneous population, with low albedo objects. The average value for the geometric albedo is 0.07$\pm$0.03, which is slightly larger than the 0.040$\pm$0.005 found by \citet{Ted89}. Larger geometric albedos, although $<$20\% and possibly due to large uncertainties, can be encountered \citep{Gra11}. Some extremely low albedos have also been reported, such as for (1173)~Anchises which displays one of the lowest: 2.7$\pm$0.7\% \citep{Hor12}. Several values are therefore considered for the albedo of Trojans, in order to study the influence of this parameter: we show results for a Bond albedo of 2\%, 5\%, 10\% and 20\% (i.e. geometric albedo of 5\%, 13\%, 26\%, and 51\%).

\paragraph{Dynamical evolution}
The dynamical evolution of Trojans can strongly affect their thermal history. Because the survival of a primordial Jupiter Trojan population remains uncertain, we consider that Trojans were implanted into their current orbits from the outer solar system, between 5 and 30~AU. \citet{Nes13} suggest that in the framework of the ``jumping-Jupiter'' model, about half of the current Trojan population would have spent between 1,000 and 10,000~yrs with perihelion distances ranging from 1.5 to 3~AU. We therefore consider two cases of dynamical evolution, one for Trojans which never orbited closer to the Sun than 5~AU, and one for Trojans which may have orbited closer to the Sun, following the time and perihelion distance constraints suggested by \citet{Nes13}.

\section{Results}

\subsection{Surface peak temperature}

\begin{figure}
\begin{center}
\includegraphics[width=\linewidth]{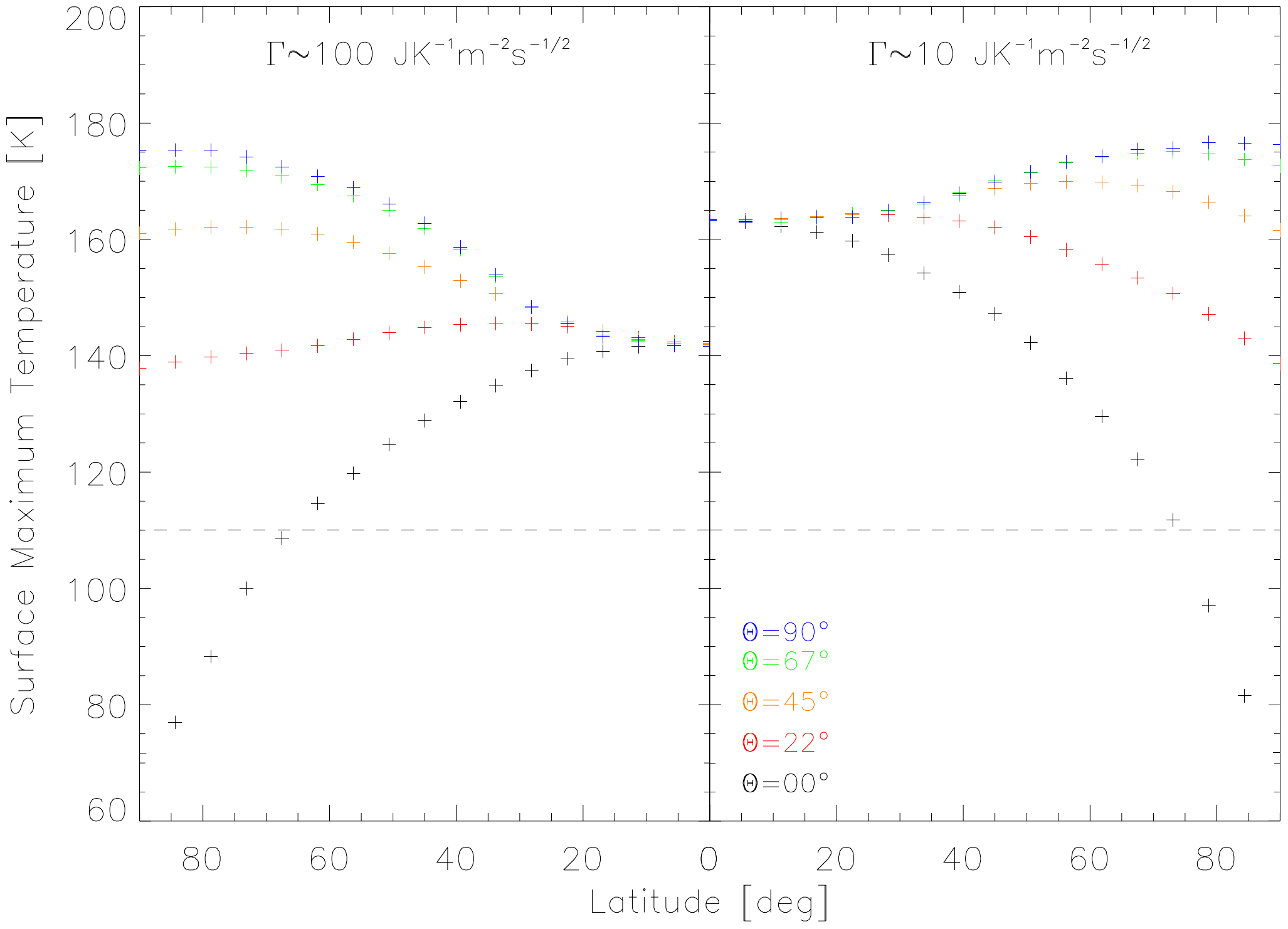}
\caption{\label{obliquity} Maximum temperature at the surface of Trojans as a function of obliquity: 0$^{\circ}$, 22$^{\circ}$, 45$^{\circ}$,67$^{\circ}$ and 90$^{\circ}$, for a 2\% Bond albedo. Left: high thermal inertia, right: low thermal inertia. The dashed line highlights the temperature of 110~K, which corresponds to the limit under which surface water ice could survive against sublimation for the age of the solar system.} 
\end{center} 
\end{figure}

\begin{figure}
\begin{center}
\includegraphics[width=\linewidth]{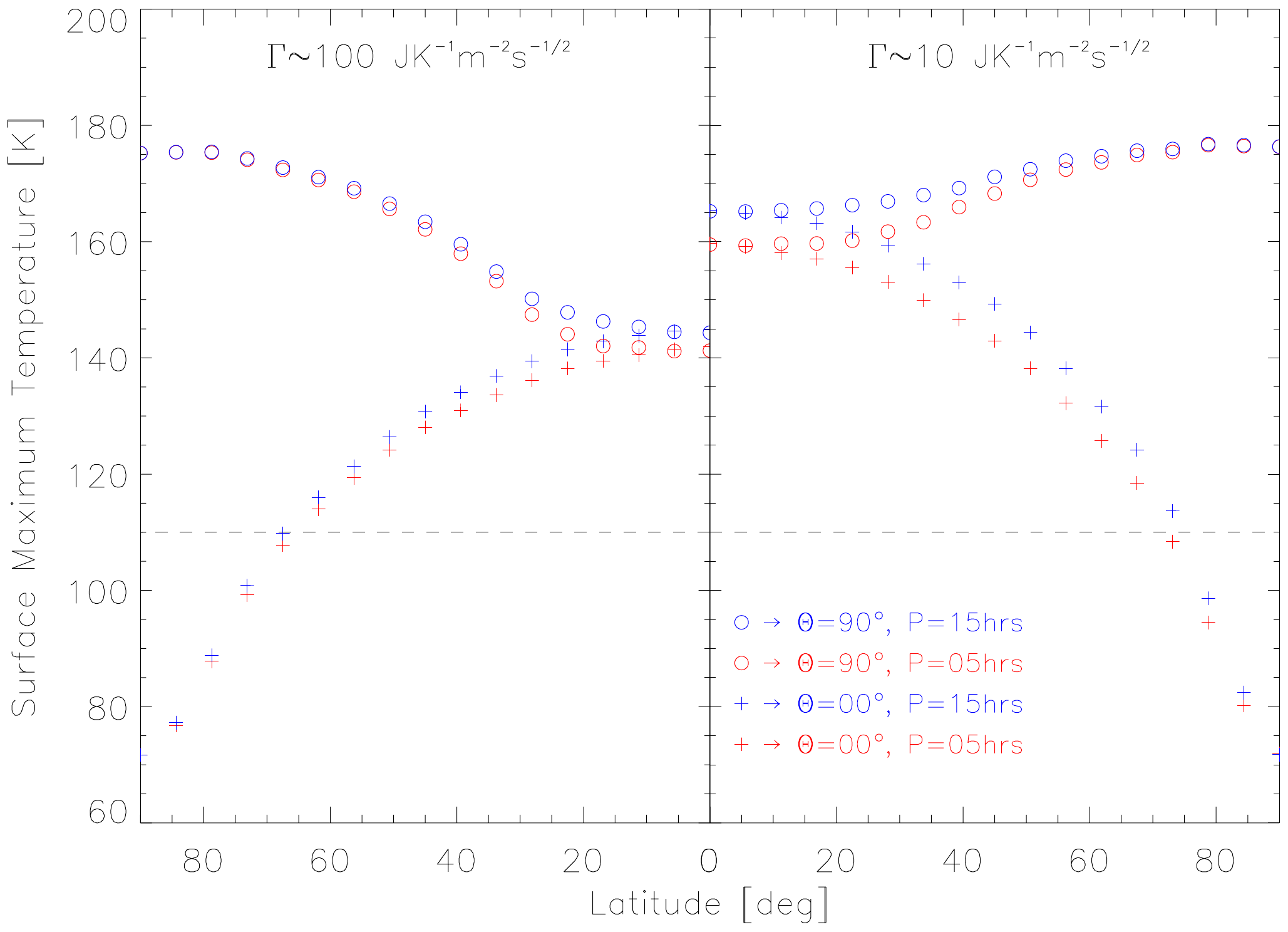}
\caption{\label{rotation} Maximum temperature at the surface of Trojans as a function of rotation period (5 and 15~hrs) and obliquity (0$^{\circ}$ and 90$^{\circ}$), for a 2\% Bond albedo. Left: high thermal inertia, right: low thermal inertia. The values for our reference rotation period (10~hrs) are not displayed for clarity, but fall between those for 5 and 15~hrs. The dashed line highlights the temperature of 110~K, which corresponds to the limit under which surface water ice could survive against sublimation for the age of the solar system.} 
\end{center} 
\end{figure}

\begin{figure}
\begin{center}
\includegraphics[width=\linewidth]{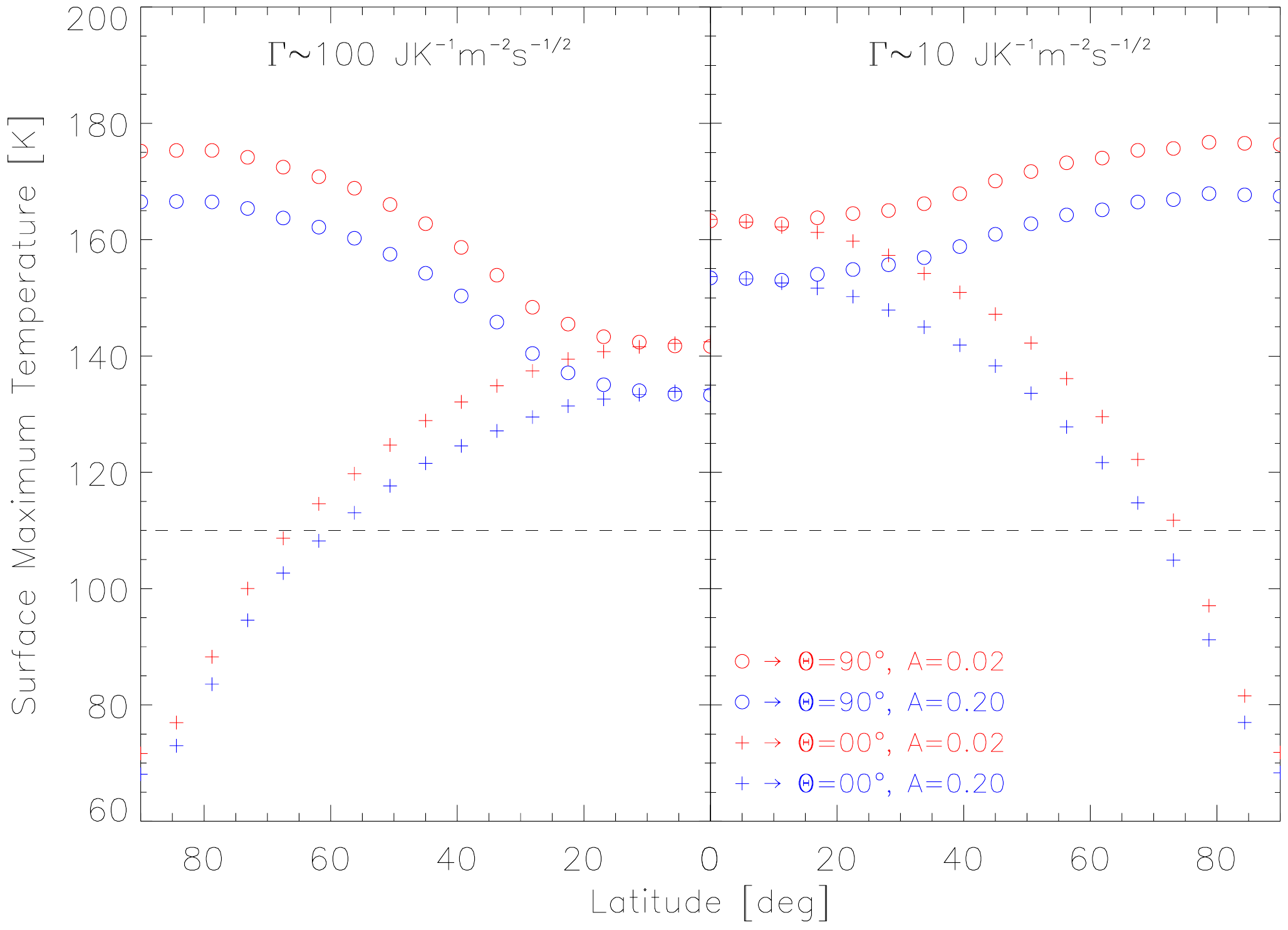}
\caption{\label{compplot} Maximum temperature at the surface of Trojans as a function of obliquity (0$^{\circ}$ and 90$^{\circ}$) and Bond albedo (2\% and 20\%). Left: high thermal inertia, right: low thermal inertia. The dashed line highlights the temperature of 110~K, which corresponds to the limit under which surface water ice could survive against sublimation for the age of the solar system.} 
\end{center} 
\end{figure}

\begin{figure}
\begin{center}
\includegraphics[width=\linewidth]{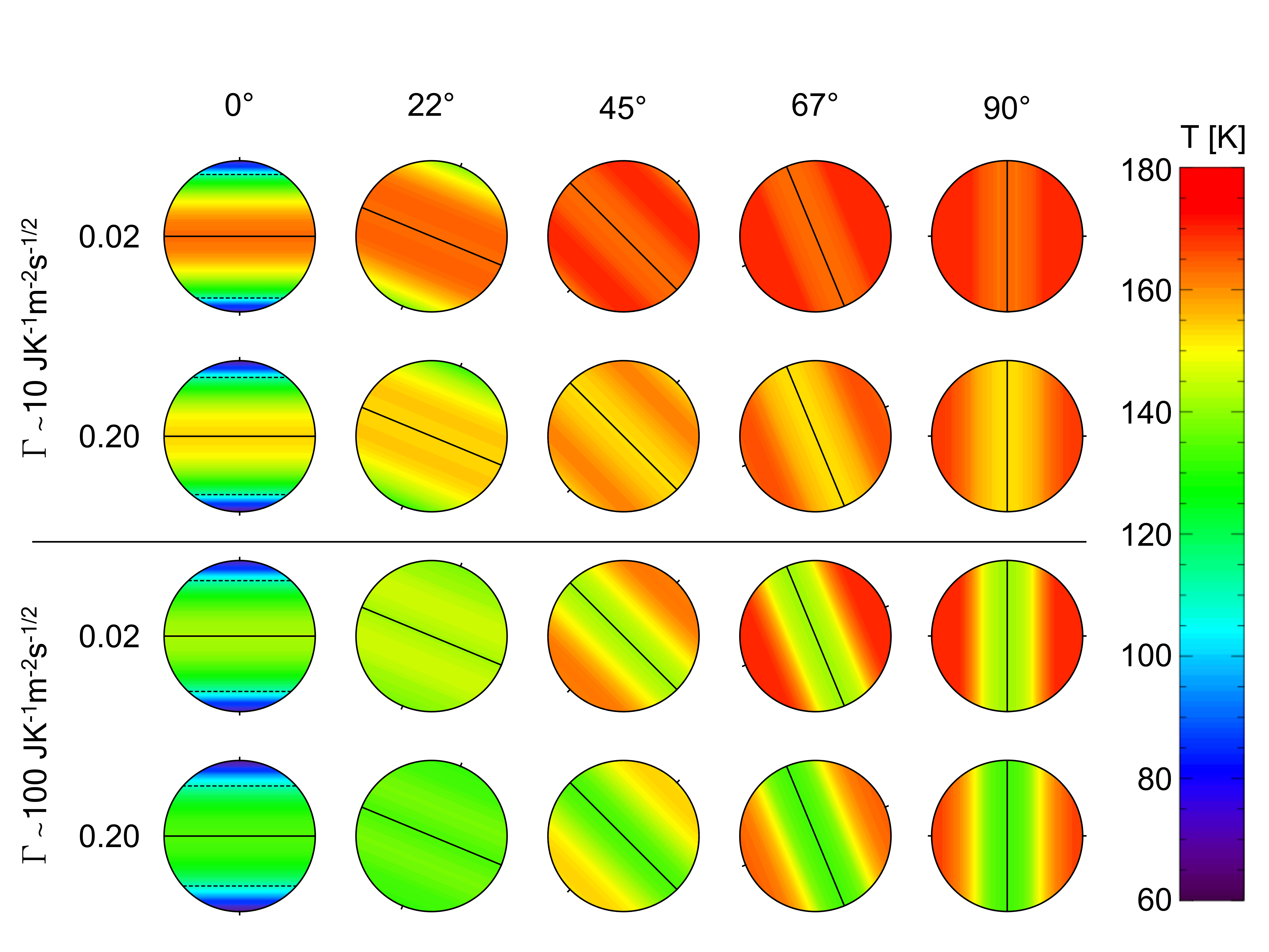}
\caption{\label{comparison} Distribution of the peak temperature at the surface of Trojans, as a function of Bond albedo (2\% and 20\%) and obliquity (0$^{\circ}$ to 90$^{\circ}$), for a low thermal inertia (top panel) and high thermal inertia (bottom panel). Dashed lines highlight the 110~K limit under which surface water ice could survive against sublimation for the age of the solar system.} 
\end{center} 
\end{figure}

Because the only heat source considered in this work is insolation, the surface is the place where the strongest temperature variations occur. It is therefore a good place to explore the influence of physical and orbital parameters relevant to our problem. In this section, we consider the simple case of Trojans with $a$=5~AU. 
The stability of ice at, or close to, the surface depends on the annual peak temperature as a function of latitude. Figure \ref{obliquity} shows this peak temperature for the various obliquities and thermal inertias we are testing (the reference rotation period of 10~hrs is considered). Because the Trojan orbit we are considering in this work is circular, the peak temperature is symmetrical with respect to the object equator. We therefore show the peak temperature for only one hemisphere. For $\Theta>20^{\circ}$, the latitudinal variations of the subsolar point induce an efficient heating of the surface at all latitude, which induces smaller variations of the peak temperature with latitude than for $\Theta<20^{\circ}$. The influence of the rotation period and albedo is shown in Fig.\ref{rotation} and \ref{compplot} respectively. The rotation period has a limited effect on the peak temperature, with variations of $\sim$5~K for the equator between the two extreme cases with rotation periods of 5~hrs and 15~hrs. There is a$~\sim$10~K difference in peak temperature between a 0.02 and a 0.2 Bond albedo. These results are summarized in Fig.\ref{comparison}, where the peak temperature distributions are given as a function of the most influential parameters: obliquity, thermal inertia, and albedo.

Below 110~K, water ice can be stable at the surface of Trojans for the age of the solar system. This limit is therefore shown as a dashed line on Fig.\ref{obliquity} to \ref{comparison}. This highlights the fact that some areas of a Trojan surface could remain cold enough to sustain water ice, like the polar regions in our simulations for $\Theta=0^{\circ}$. However, we need to emphasize the fact that the cometary past of Trojans was not studied here, and a specific modeling would be required in order to assess the actual survival of water ice at the surface of these objects. In addition, Trojans could have orbited closer to the Sun than 5~AU \citep[e.g. in the framework of the ``jumping-Jupiter'' model,][]{Nes13}, which could have induced a significant loss of water ice. They might also have sustained variations of their spin axis due to impacts and outgassing, which might have also limited the actual stability of water ice at their surface. Within these limitations, we find that $\sim$10\% of water ice could have survived, at most, which would make it very difficult to detect from ground-based spectroscopic observations. 

\subsection{\label{5au}Crust thickness at 5~AU}

\begin{figure}
\begin{center}
\includegraphics[width=\linewidth]{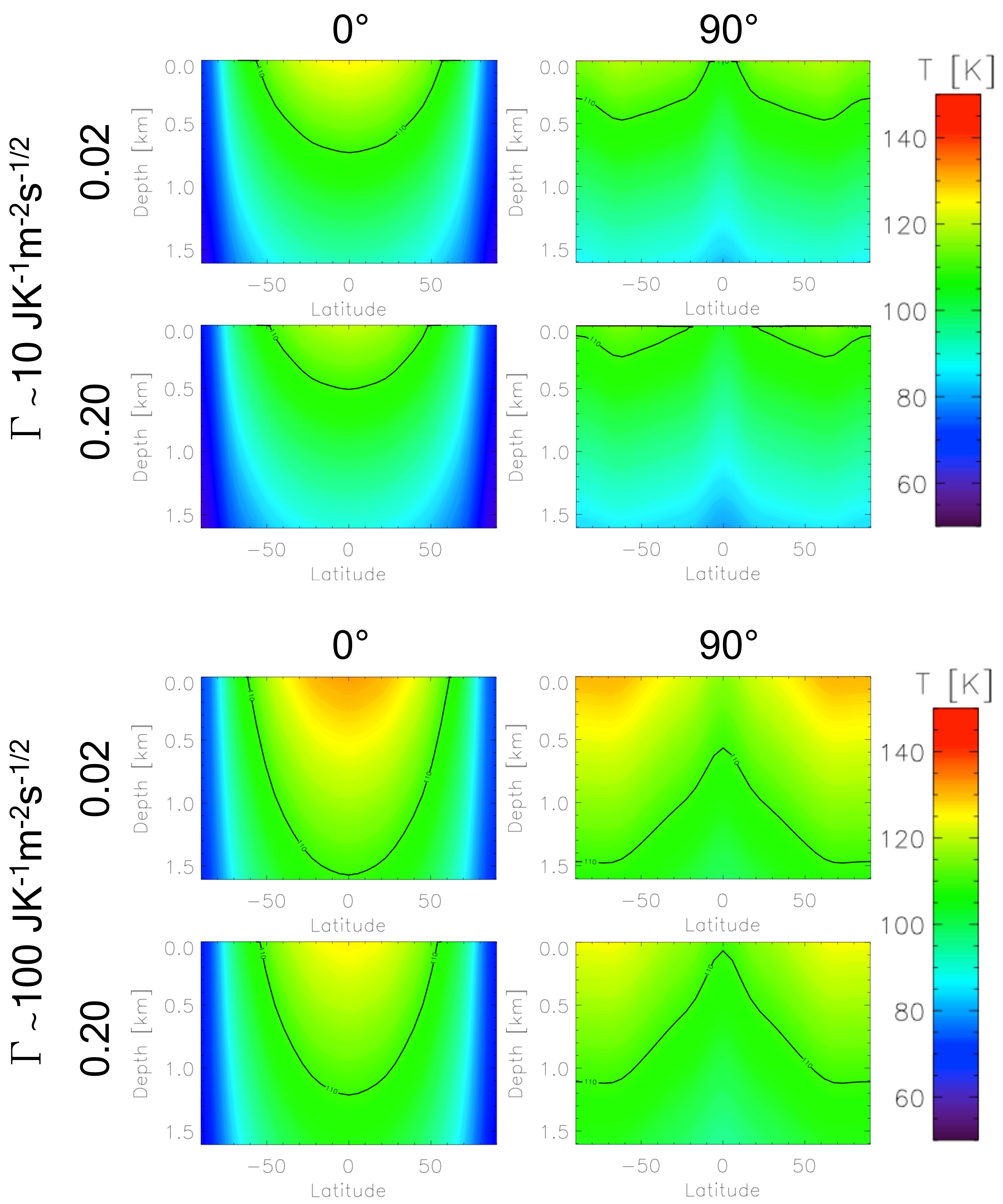}
\caption{\label{intdistrib} Temperature distribution inside the dusty crust, as a function of Bond albedo (2\% and 20\%) and obliquity (0$^{\circ}$ and 90$^{\circ}$), for a low thermal inertia (top panel) and high thermal inertia (bottom panel). The solid line shows the 110~K isotherm.} 
\end{center} 
\end{figure}

If Trojans never orbited closer to the Sun than 5~AU, their thermal history would be dominated by present-day heating conditions, which are more intense than any conditions sustained at larger heliocentric distances. Therefore, by studying the stability of ice in subsurface layers at 5~AU for the age of the solar system, we can obtain an upper limit of the crust thickness required for ice to survive underneath, for the $\sim$50\% of Trojans which may have been directly implanted into their current orbits, without orbiting closer to the Sun.
At sufficient depth, typically below the orbital skin depth,
the material is not sensitive to the surface peak temperature, but to the mean annual temperature.
The propagation of the heat wave inside the porous dusty crust, from an initial internal temperature of 50K, is shown in Fig.\ref{intdistrib}, as a function of thermal inertia, albedo and obliquity. 

\begin{figure}
\begin{center}
\includegraphics[width=\linewidth]{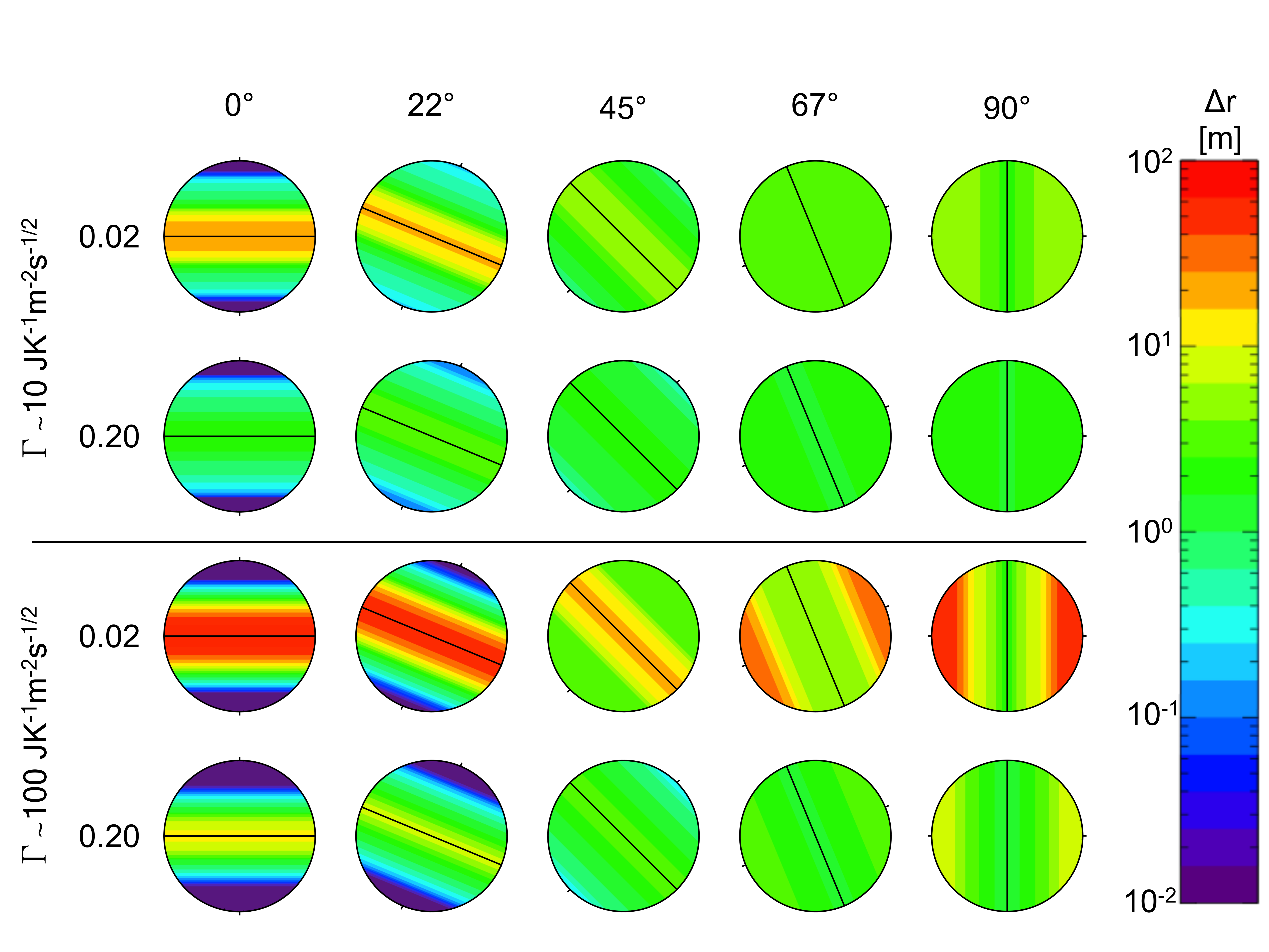}
\caption{\label{croute} Thickness of the porous dusty crust at the surface of Trojans, under which water ice could have survived for the age of the solar system. It is shown as a function of as a function of Bond albedo (2\% and 20\%) and obliquity (0$^{\circ}$ to 90$^{\circ}$), for a low thermal inertia (top panel) and high thermal inertia (bottom panel).} 
\end{center} 
\end{figure}

Using Eq.(\ref{jint}), we can directly constrain the thickness of the porous dusty crust required for water ice to survive, from the temperature distribution inside the crust. The results are given in Fig.\ref{croute}. For the low inertia cases, water ice can survive under a 5-m thick crust in most cases, i.e. except for a 2\% albedo and $\Theta=0^{\circ}$ and $22^{\circ}$. A high thermal inertia and a 20\% albedo are also producing favorable surviving conditions for water ice, since it can be found below 10~m. Our most extreme cases are for a high thermal inertia and low albedo (2\%), where most of the incoming heat is transported toward the subsurface layers. In these cases, the crust thickness can reach 85~m. 

Our modeling assumes some thermo-physical properties for Trojans, which might influence the real depth at which water ice could survive. For example, we assumed completely uniform thermo-physical parameters throughout the dusty crust. Local variations of the porosity, thermal inertia or albedo could in fact induce variations of the crust thickness. 
The fact that Trojans are not spheres should not change the basic conclusions. The internal structures that we describe here with simple concentric spherical interfaces would be determined by isotherms defined by the real shape of the objects. 

\subsection{\label{close}Crust thickness after an orbital evolution at small heliocentric distances}

We now consider the case of Trojans which might have orbited closer to the Sun. Although the period of orbital evolution close to the Sun suggested by \citet{Nes13} is very short compared to the age of the solar system, the heating conditions sustained during this phase might dominate the Trojan final stratigraphy. Although outgassing and dust mantling are not specifically studied in this work, we find it interesting to provide a direct comparison between the crust thicknesses found in section \ref{5au}, with those which could be expected (within our modeling limits) at the end of this phase of intense cometary activity.
In order to be conservative, we consider orbits with $a$=1.5~AU and $a$=3~AU, and study the stability of ice in subsurface layers after 1,000 and 10,000~yrs. 

\begin{figure}
\begin{center}
\includegraphics[width=\linewidth]{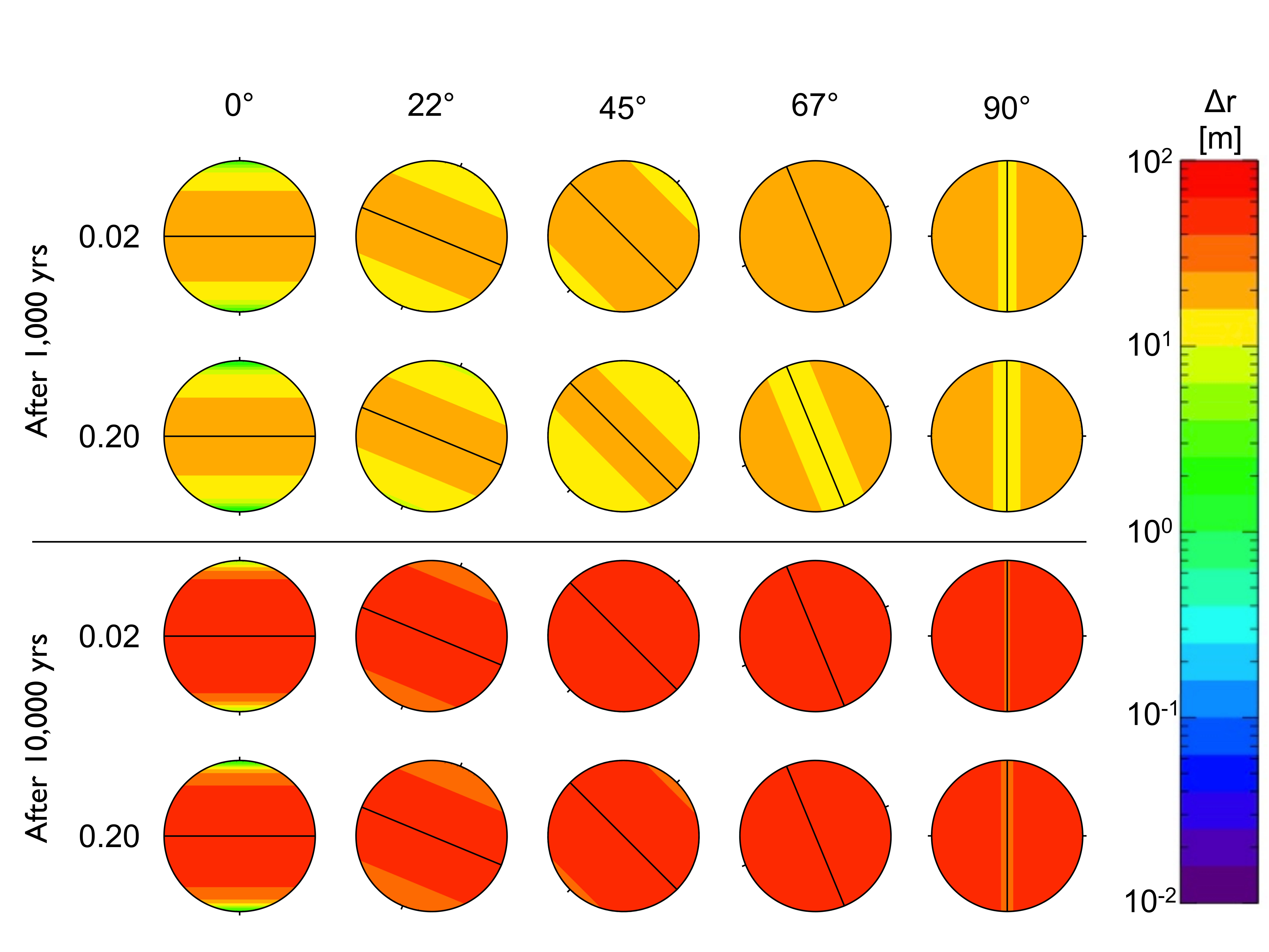}
\caption{\label{croute15} Thickness of the porous dusty crust at the surface of Trojans, under which water ice could have survived after 1,000~yrs (top) and 10,000~yrs (bottom) of evolution at 1.5~AU. It is shown as a function of Bond albedo (2\% and 20\%) and obliquity (0$^{\circ}$ to 90$^{\circ}$), for a low thermal inertia.} 
\end{center} 
\end{figure}

\begin{figure}
\begin{center}
\includegraphics[width=\linewidth]{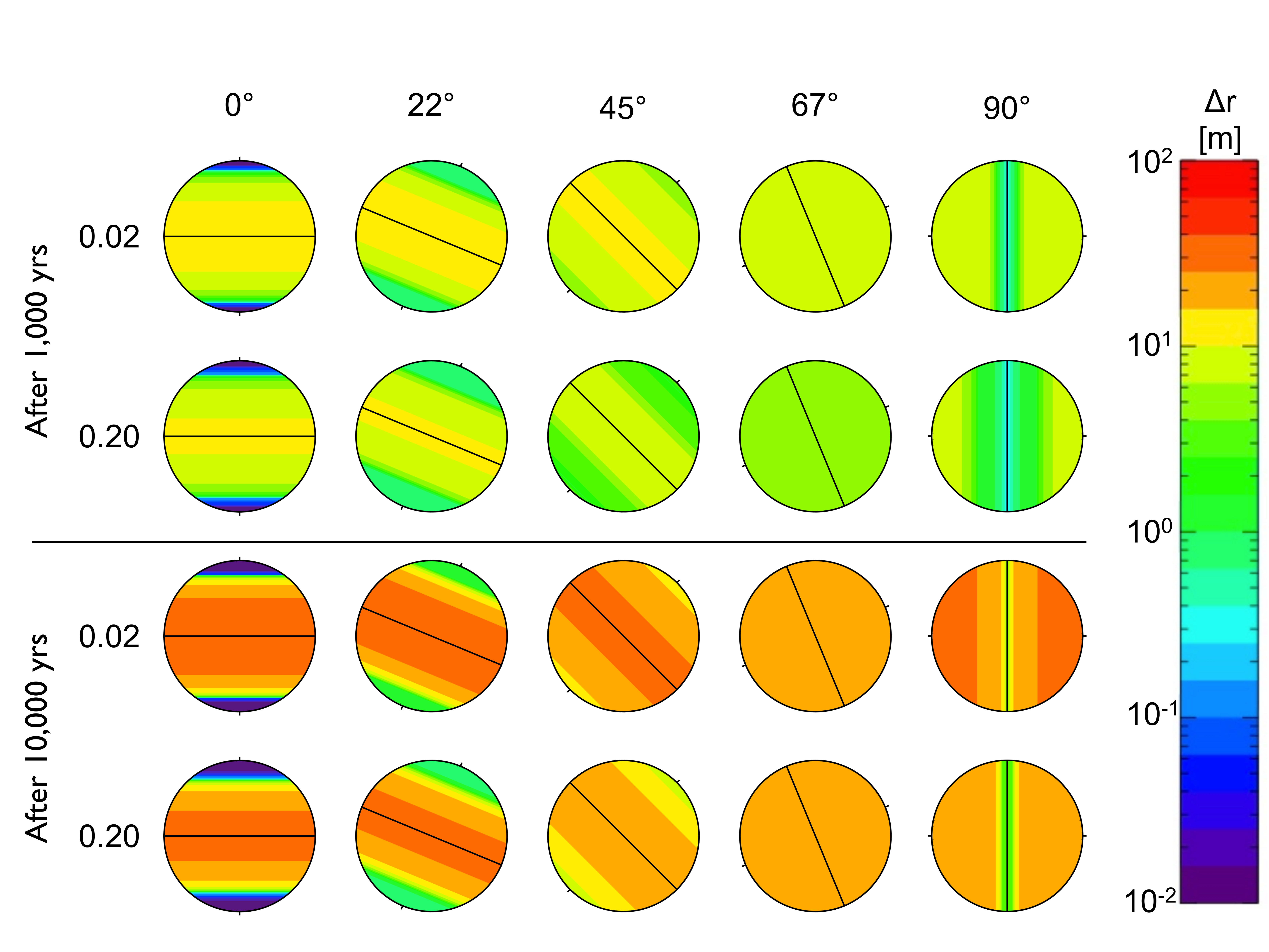}
\caption{\label{croute30} Thickness of the porous dusty crust at the surface of Trojans, under which water ice could have survived after 1,000~yrs (top) and 10,000~yrs (bottom) of evolution at 3~AU. It is shown as a function of Bond albedo (2\% and 20\%) and obliquity (0$^{\circ}$ to 90$^{\circ}$), for a low thermal inertia.} 
\end{center} 
\end{figure}

The results for a low thermal inertia are given on Fig.\ref{croute15} and \ref{croute30}. They confirm that the evolution of the temperature distribution inside Trojans, and their associate internal structure, would be dominated by this stage of orbital evolution close to the Sun. We find that stable water ice would be found 30~m below the surface after 10,000~yrs at 3~AU (50~m for a high thermal inertia), and 50~m below the surface after 10,000~yrs at 1.5~AU (80~m for a high thermal inertia). Following this period of orbital evolution close to the Sun, these Trojans may have been implanted in the Lagrangian clouds, where less intense heating conditions would not change this internal stratigraphy, unless surface layers were progressively eroded by impacts for example. Qualitatively, we can mention that eroded or locally blown-off crusts could bring some fresh ice close to the surface, so that the whole process of dust mantle formation would start again, until a new layer of sufficient thickness was formed to protect the ice underneath. Heating conditions at the current location of Trojans should then prevail, and associated crust thicknesses would be those presented in section \ref{5au}.

\section{Discussion and conclusion}

We have studied the survival of buried water ice inside Jupiter Trojans. We have used a simple description of gas flow through a porous dusty layer, coupled with a complex model of the thermal evolution of these objects, so as to provide constraints on the thickness of the dust mantle which is required for ice to survive underneath. Because such a porous dusty crust is a very good insulator, we find that water ice could survive in close subsurface layers over the age of the solar system: it can be found on the top 100~m below the surface in all the cases we have investigated, even for Trojans which might have orbited very close to the Sun during their dynamical evolution. 
It is worth comparing these results with the works of \citet{Sch08} and \citet{Pri09}, who studied the stability of water ice in the Main Belt of asteroids. \citet{Sch08} used a similar simple description of the gas flow through a porous dusty mantle, and found that water ice would be able to survive in the Main Belt on the top few meters over the age of the solar system. Using a more complex model, including ice phase transitions and dust mantling process, \citet{Pri09} found instead that water ice should be stable between $\sim$50 to 150~m below the surface. Since we find that water ice should be stable $\sim$10m below the surface (on average) after 10,000~yrs of evolution at 3~AU, we can argue that our results fall in between those two different modeling results. Indeed, we find that water ice can be sustained 5~m below the surface in most cases at 5~AU, and $\sim$1~m below the surface on average. 

To test the volatile-rich nature of Trojans, we would need to be able to detect the ice, if present. We could additionally want to test the models of the solar system evolution, for example by distinguishing between two possible subpopulations of Trojans, relative to the depth at which water ice is present. Given the depths described above, the question arises whether such a study would be possible, by radar observations for example. Indeed, as an electromagnetic wave propagates through a medium, it is attenuated, scattered and dispersed. It thus carries information on the objects' internal composition and structure. In particular, a change in composition causes a discontinuity of the dielectric constant (between the dusty superficial layer and the ice/dust mixture in the core in our case). Such a discontinuity can be detected by a radar sounder. Direct probing of subsurfaces down to km depths is possible by using low frequency sounding radars \citep[see for example][]{Kof10}. High frequency radars give access to shallow subsurface layers, but with a higher vertical resolution. For example, \citet{Tho06} showed that high frequency radars could probe 10 to 50~m through the lunar regolith. Therefore, it is reasonable to expect that in situ radar observations would be able to constrain the actual presence or absence of water ice in subsurface layers of Trojans, and therefore bring crucial insights on the solar system evolution.


\paragraph{Acknowledgements}
This work was supported by a NASA Herschel grant to David Jewitt, and the ESA fellowship program. We thank David Jewitt and Pierre Vernazza for insightful discussions, Joshua Emery and an anonymous referee for comments which significantly improved the manuscript. 


\clearpage

\clearpage

\clearpage


\clearpage


\clearpage

\clearpage

\clearpage

\clearpage

\end{document}